\documentclass{PoS}

\title{NRQCD based S- and P-wave bottomonium spectra 
at finite temperature from $48^3 \times 12$ lattices with $N_f=2+1$ light HISQ flavors}

\ShortTitle{NRQCD based S- and P-wave Bottomonium spectra
at finite temperature}

\author{Seyong Kim\\
Department of Physics, Sejong University, Seoul 143-747, Korea\\
  E-mail: \email{skim@sejong.ac.kr}}

\author{\speaker{Peter Petreczky}\\
        Physics Department, Brookhaven National Laboratory,
  Upton, NY 11973, USA\\
        E-mail: \email{petreczk@quark.phy.bnl.gov}}

\author{Alexander Rothkopf\\
        Institute for Theoretical Physics,  Heidelberg
  University, Philosophenweg 16, 69120 Heidelberg, Germany\\
        E-mail: \email{rothkopf@thphys.uni-heidelberg.de}}

\abstract{We study S-wave and P-wave bottomonium spectral functions
at non-zero temperature in 2+1 flavor QCD using the NRQCD formulation
for bottom quarks. We use a novel Bayesian approach to reconstruct 
the spectral functions and find that  $\chi_{b1}$ survives up to
$T=249 {\rm MeV}$. We also study the effect of different temporal
discretizations of the NRQCD formalism on the bottomonium
correlation functions.}

\FullConference{The 32nd International Symposium on Lattice Field Theory,\\
		23-28 June, 2014\\
		Columbia University New York, NY}

\begin{document}

\section{Introduction}
Quarkonium suppression has been suggested as the signal for
the formation of a deconfined medium in heavy ion collisions
by Matsui and Satz \cite{Matsui:1986dk}. Since this seminal work
the properties of quarkonium at non-zero temperature became a
subject of intense theoretical studies (see e.g. Ref. \cite{Mocsy:2013syh} for a recent review ).
In recent years interesting experimental results appeared on
bottomonium suppression in heavy ion collisions
\cite{Chatrchyan:2012lxa,Adamczyk:2013poh,Abelev:2014nua} 
making first principle lattice QCD studies of bottomonium properties at non-zero
temperature timely. The study of bottomonium in lattice QCD is challenging
due to the large bottom quark mass $M_b$. In standard lattice formulations
discretization effects will be proportional to $a M_b$, and thus will
be very large for typical values of the lattice spacing $a$ available
in today's calculations. Fortunately one can use the effective field
theory approach and integrate out the physics related to the
large bottom quark mass \cite{Thacker:1990bm,Lepage:1992tx}.
This approach is known as non-relativistic QCD or NRQCD for short. It is widely used today
to study bottomonium properties at zero temperature in lattice QCD ( see e.g. Ref. \cite{Gray:2005ur} ). 
NRQCD was first proposed to study quarkonium properties at non-zero temperature in Ref. \cite{Fingberg:1997qd} 
and recently it was used for bottomonium at non-zero temperature in 2-flavor and 2+1 flavor QCD
\cite{Aarts:2010ek,Aarts:2011sm,Aarts:2012ka,Aarts:2013kaa,Aarts:2014cda} at un-physically large u/d quark
mass corresponding to pion masses of about $400$ MeV. In this contribution we will
present a study of bottomonium correlation functions and spectral functions at non-zero temperature
in 2+1 flavor QCD with nearly physical values of the strange and light quark masses using
the $48^3 \times 12$ gauge configuration generated by HotQCD collaboration \cite{Bazavov:2011nk}. 
The chiral transition temperature corresponding to these lattices is about $159$ MeV \cite{Bazavov:2011nk}. 
A more detailed account of this work can be found in Ref. \cite{Kim:2014iga}, while preliminary
results have been presented at Lattice 2013 conference \cite{Kim:2013seh}.

\section{Bottomonium in Lattice NRQCD}

In order to investigate the properties of bottomonium in a thermal
medium, we compute the correlators of heavy quarkonium using a lattice
discretization of the ${\cal O} (v^4)$ NRQCD Lagrangian
\cite{Lepage:1992tx} for bottom quarks. In this approach 
the bottom quark is not part of the thermal medium and no 
temporal boundary conditions are imposed on the bottom quark. 
The bottom quark propagator is calculated as an initial value
problem in the imaginary time direction 
with a step-size determined by the lattice spacing
and the so-called Lepage parameter $n$ (see Ref. \cite{Kim:2014iga} for
a detailed discussion). We use the value $n=2$ for the Lepage parameter
which is suitable as long as $a M_b>1.5$. We also consider $n=3$ and
$4$ in our calculations and tested that our conclusions do not depend
on the choice of $n$. The use of the NRQCD formulation in the study of
quarkonium spectral functions has at least two advantages. First,
the so-called zero mode contribution \cite{Umeda:2007hy,Petreczky:2008px}
is absent in this case, making
the analysis of the correlators easier. Second, because of the absence
of periodic boundary conditions quarkonium correlators can be studied
up up twice larger separations in Euclidean time, namely, $\tau_{max}=1/T$ instead
of $\tau_{max}=1/(2 T)$ in the usual (relativistic) formulation.

\begin{figure}
\centering
 \includegraphics[scale=0.29, angle=-90]{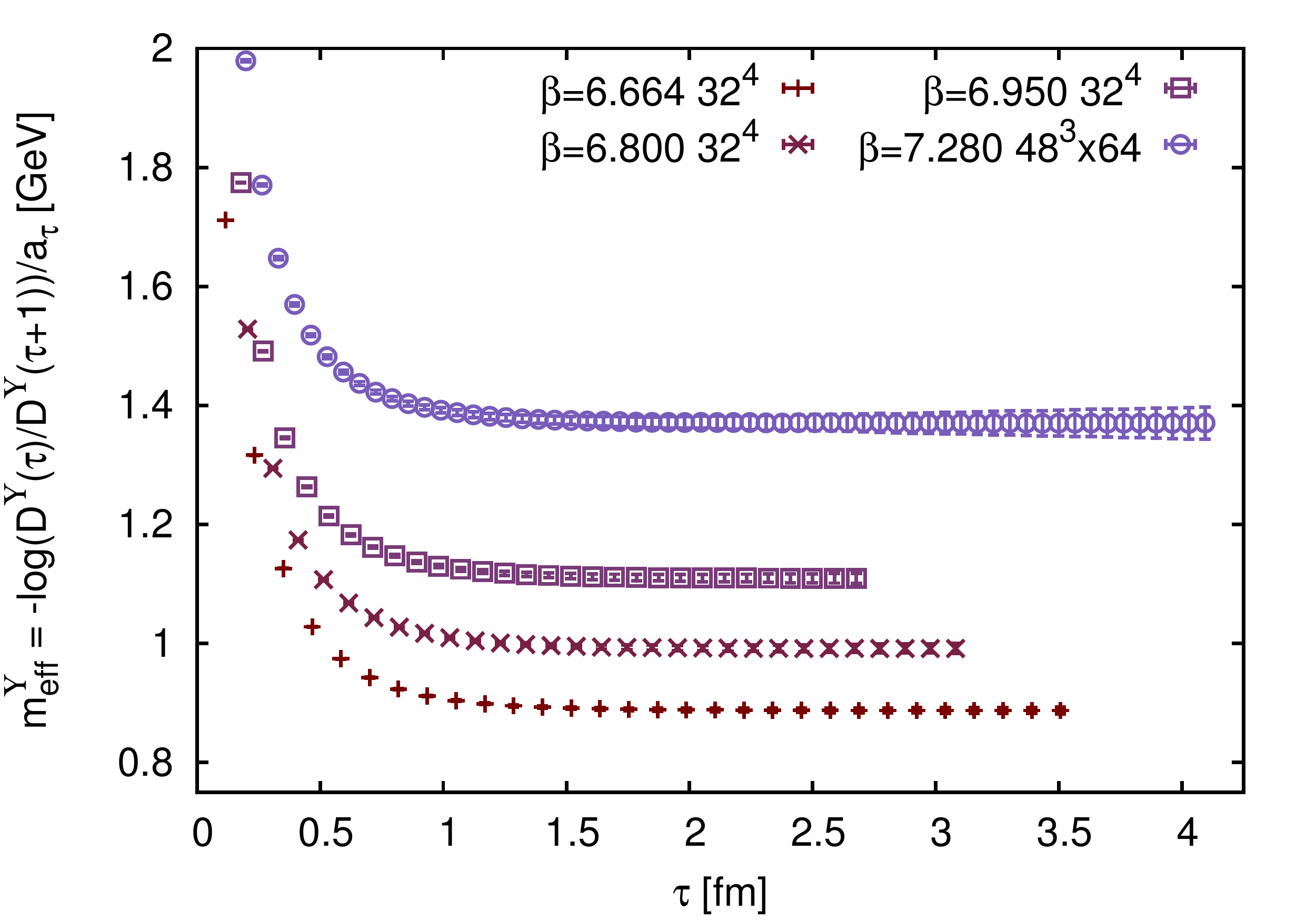}
 \includegraphics[scale=0.29, angle=-90]{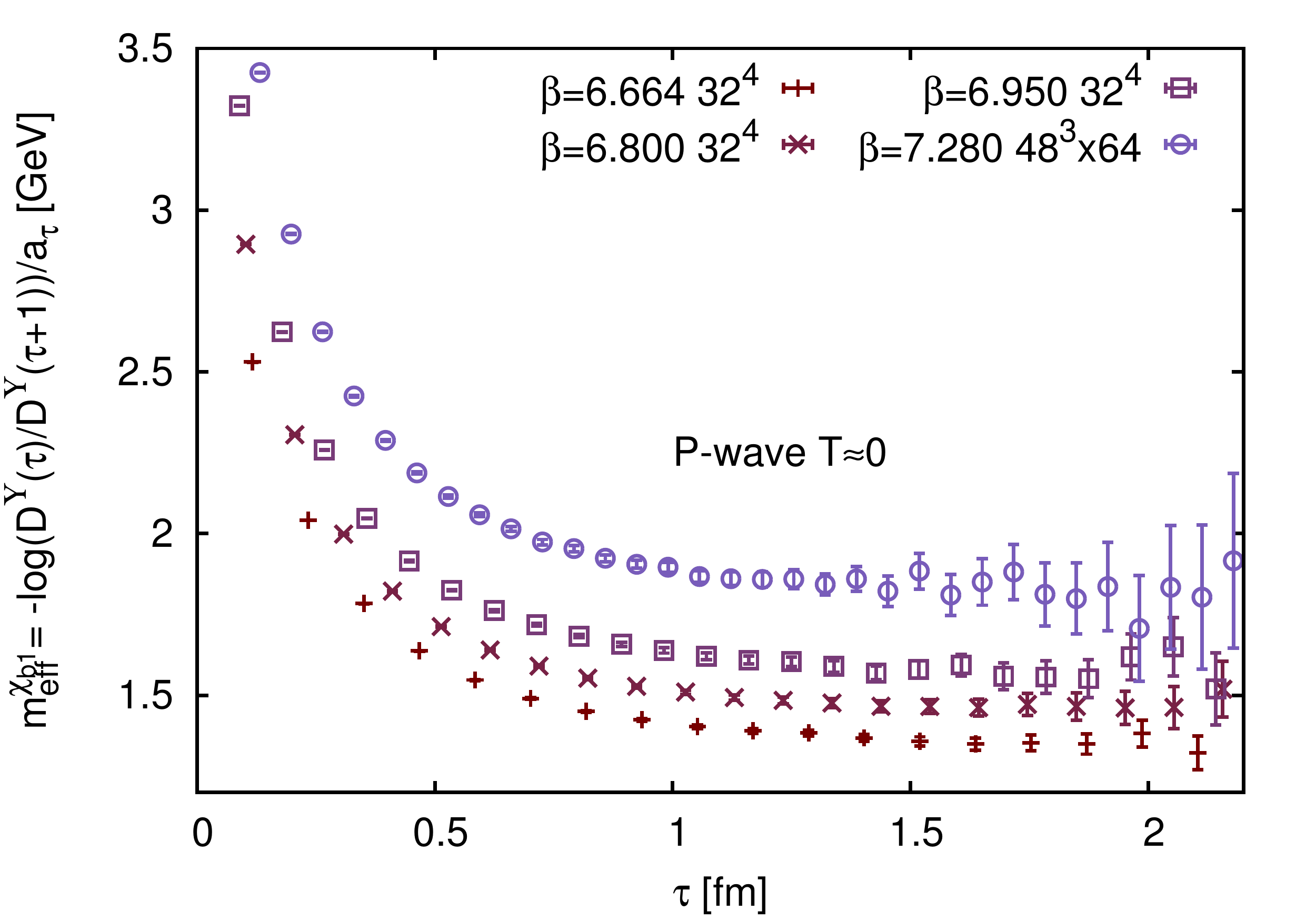}
\caption{The effective masses $m_{\rm eff}(\tau)$ in the S-wave channel (left)
  and P-wave channel (right) at $T\simeq0$ for $\beta = 6.664, 6.800,
  6.950,$ and $7.280$}
 \label{fig:effm}
\end{figure}

\begin{figure}
\centering
 \includegraphics[scale=0.28, angle=-90]{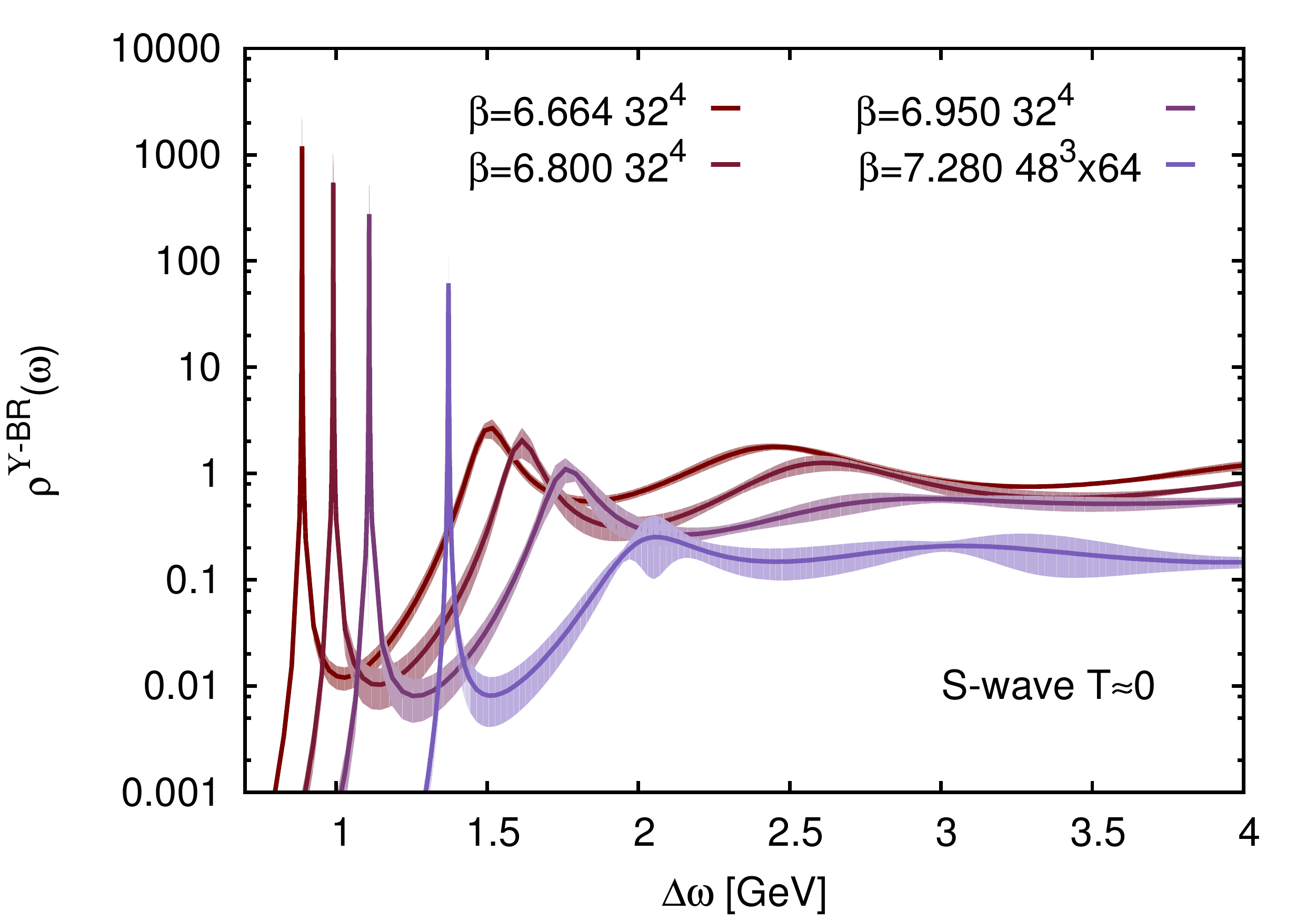}
 \includegraphics[scale=0.28, angle=-90]{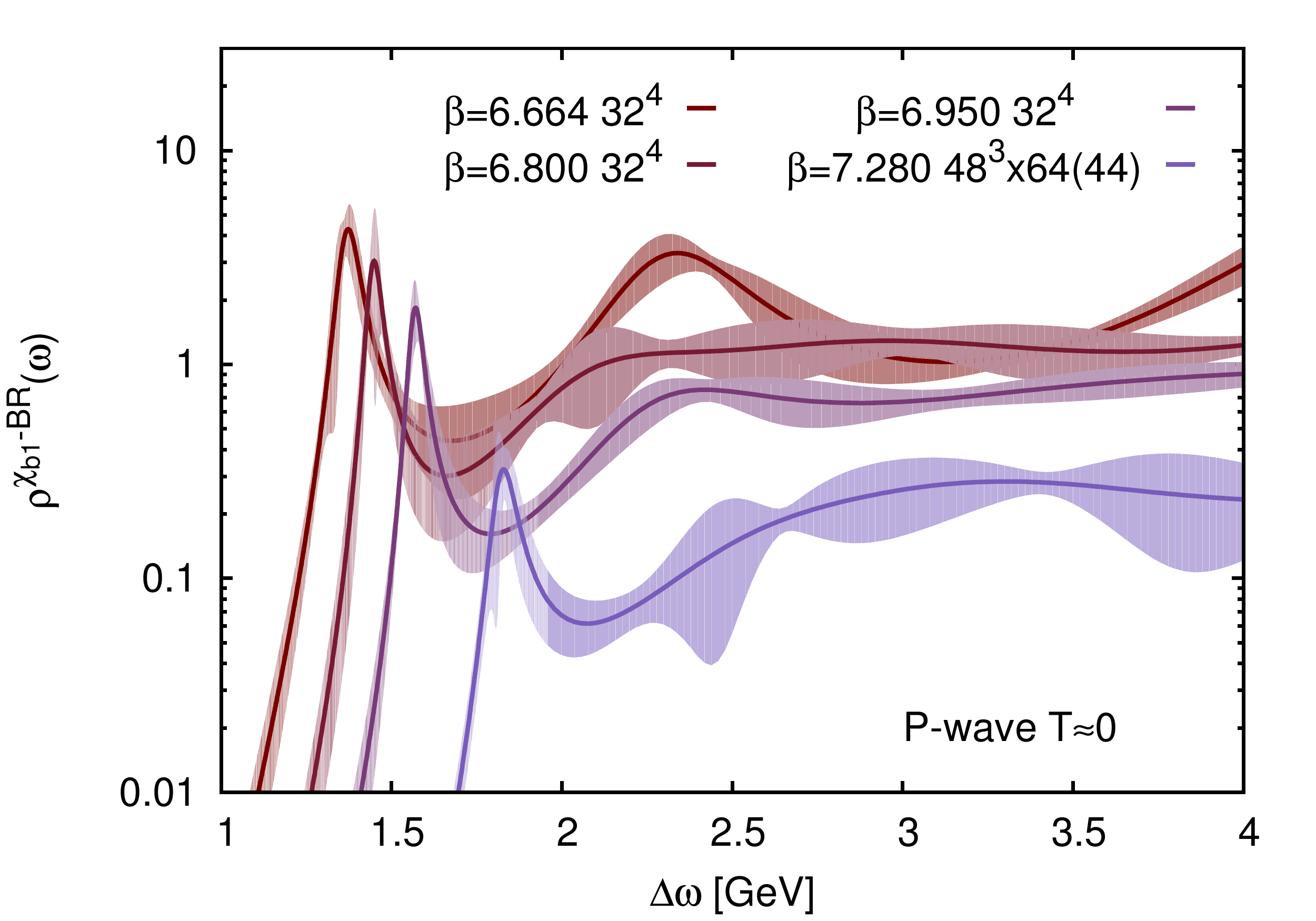}
\caption{Bayesian reconstruction of the non shifted spectra in the
  S-wave channel (left) and P-wave channel (right) at $T\simeq0$
  temperature for $\beta = 6.664, 6.800, 6.950,$ and $7.280$. Note
  that at $\beta=7.280$ we only use the first $44$ data points there
   the signal is not yet lost to round-off-errors. }
 \label{fig:spf0}
\end{figure}

We start the discussion with presenting our numerical results
at zero temperature. In Fig. \ref{fig:effm} we show the results
on the zero temperature $\Upsilon$ and $\chi_{b1}$ correlators
for four values of the gauge coupling 
$\beta=10/g^2=6.74,~6.80,~6.95$ and $7.28$ in terms of the effective
masses $a m_{eff}(\tau)=\ln\left(D(\tau/a)/D(\tau/a+1)\right)$.
These values of the gauge coupling correspond to temperatures
$T=140,~160,~184$ and $249$ MeV, respectively on $N_{\tau}=12$ lattices.
At large $\tau$ the effective mass reaches a plateau which
corresponds to the mass of the bottomonium state. 
The meson masses in NRQCD are
related to the physical meson masses by a $\beta$
dependent constant $C(\beta)$, e.g. 
\begin{equation}
M_{\Upsilon (1S)}^{\rm exp} = E^{\rm sim}_{\Upsilon (1S)} + C(\beta) ,
\end{equation}
where $E^{\rm sim}_{\Upsilon (1S)}$ is the $(1S)$ energy computed in
the $\Upsilon$ channel in NRQCD and $M_{\Upsilon}^{\rm exp}$
refers to the experimental value for the mass of the $\Upsilon$
state. Therefore, the effective masses in Fig. \ref{fig:effm} 
depend on $\beta$. 
We reconstruct the spectral function of S-wave and P-wave quarkonia using
the novel Bayesian approach described in Ref. \cite{Burnier:2013nla}. 
More precisely 
we consider $\Upsilon$ and $\chi_{b1}$ states which corresponds 
to the vector and axial-vector channels in relativistic QCD. The
results are shown in Fig. \ref{fig:spf0}. The
$\Upsilon(1S)$ state is very well determined from the reconstruction. The
second bump corresponds to excited states, mostly $\Upsilon(2S)$.
In the case of P-wave spectral function the first peak corresponding
to the $\chi_{b1}(1P)$ state is broader and the statistical errors on
the spectral functions are larger. This is due to at least two
effects. First, the mass of the $\chi_{b1}(1P)$ state is larger than
the mass of $\Upsilon(1S)$, so the signal-to-noise ratio is smaller.
Second, the relative contribution of the bound state peak to the correlator
compared to the continuum contribution is smaller than in the S-wave.
To get the spectral functions in terms of the physical energy the curves in
Fig. \ref{fig:spf0} should be shifted by the constant $C(\beta)$ defined above.

\begin{figure}
\begin{center}
\vspace{-0.4cm}
 \includegraphics[scale=0.29,angle=-90]{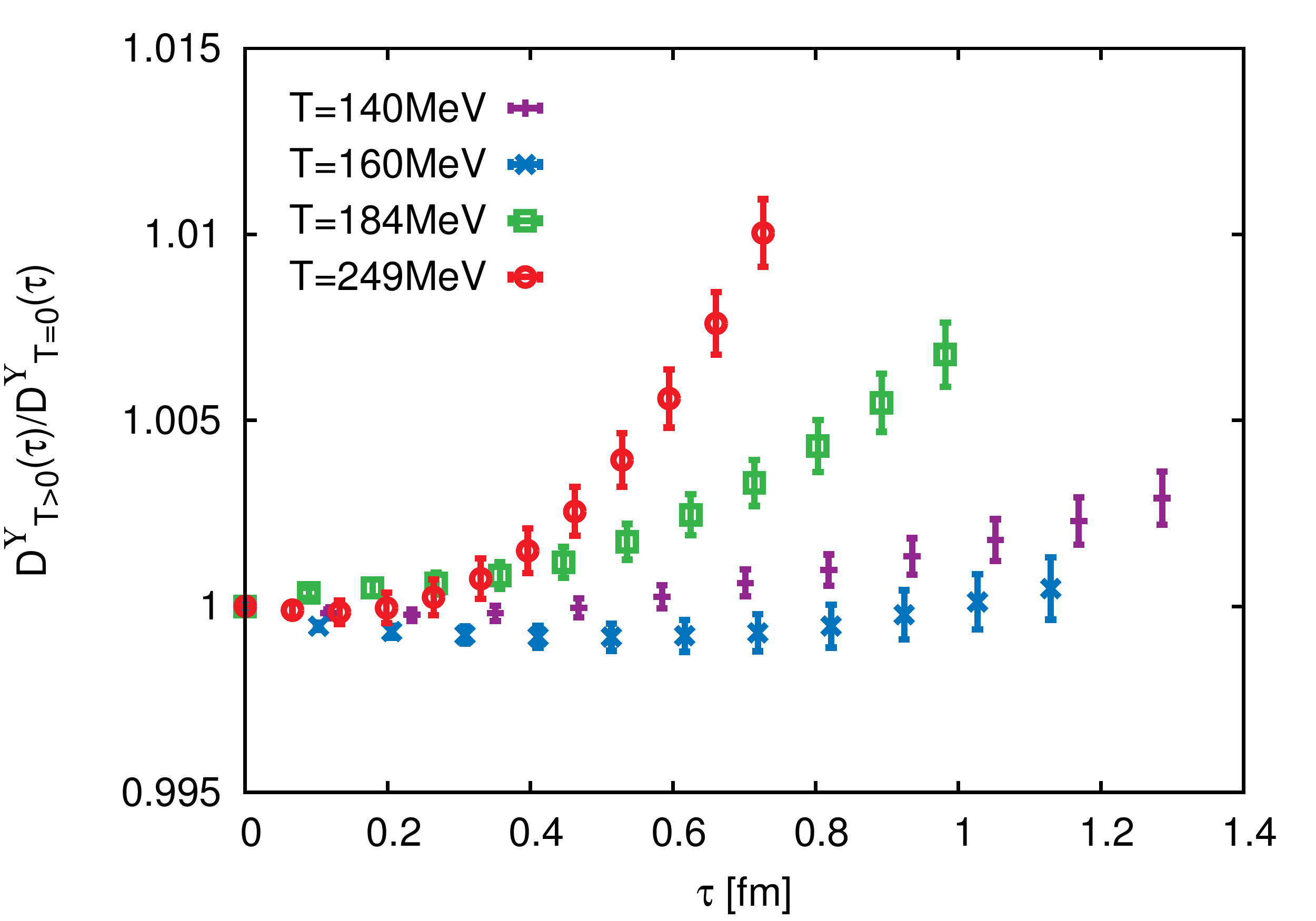}
 \includegraphics[scale=0.29,angle=-90]{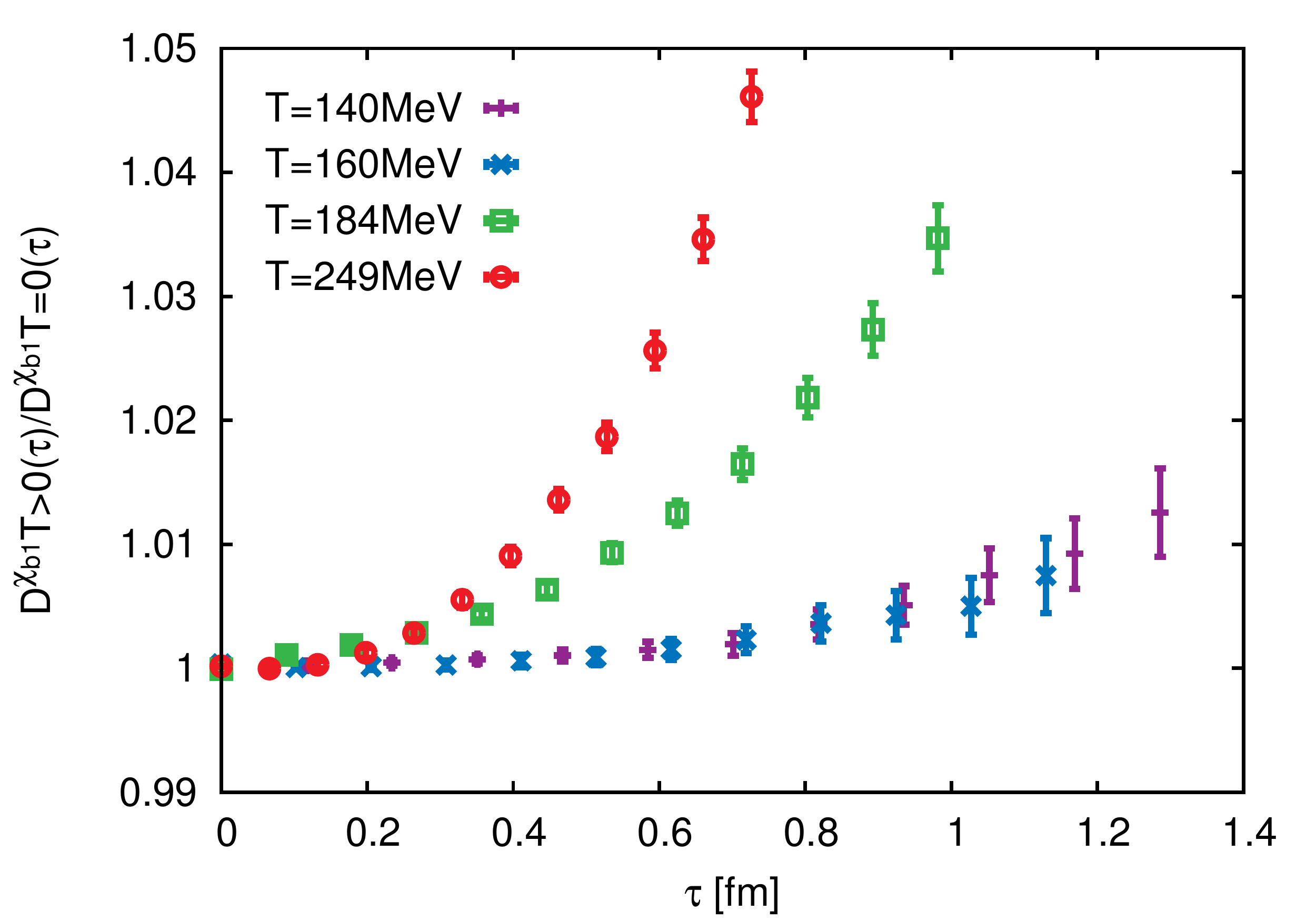}
\end{center}\vspace{-0.2cm}
\caption{Ratios of the Euclidean correlator at finite temperature and
  close to $T=0$ for the $\Upsilon$ (left) and $\chi_{b1}$ channel at
  the lattice spacings, where $T\simeq0$ measurements were carried
  out.}
 \label{fig:corr_rat}
\end{figure}

\begin{figure}
\begin{center}
\includegraphics[scale=0.29,angle=-90]{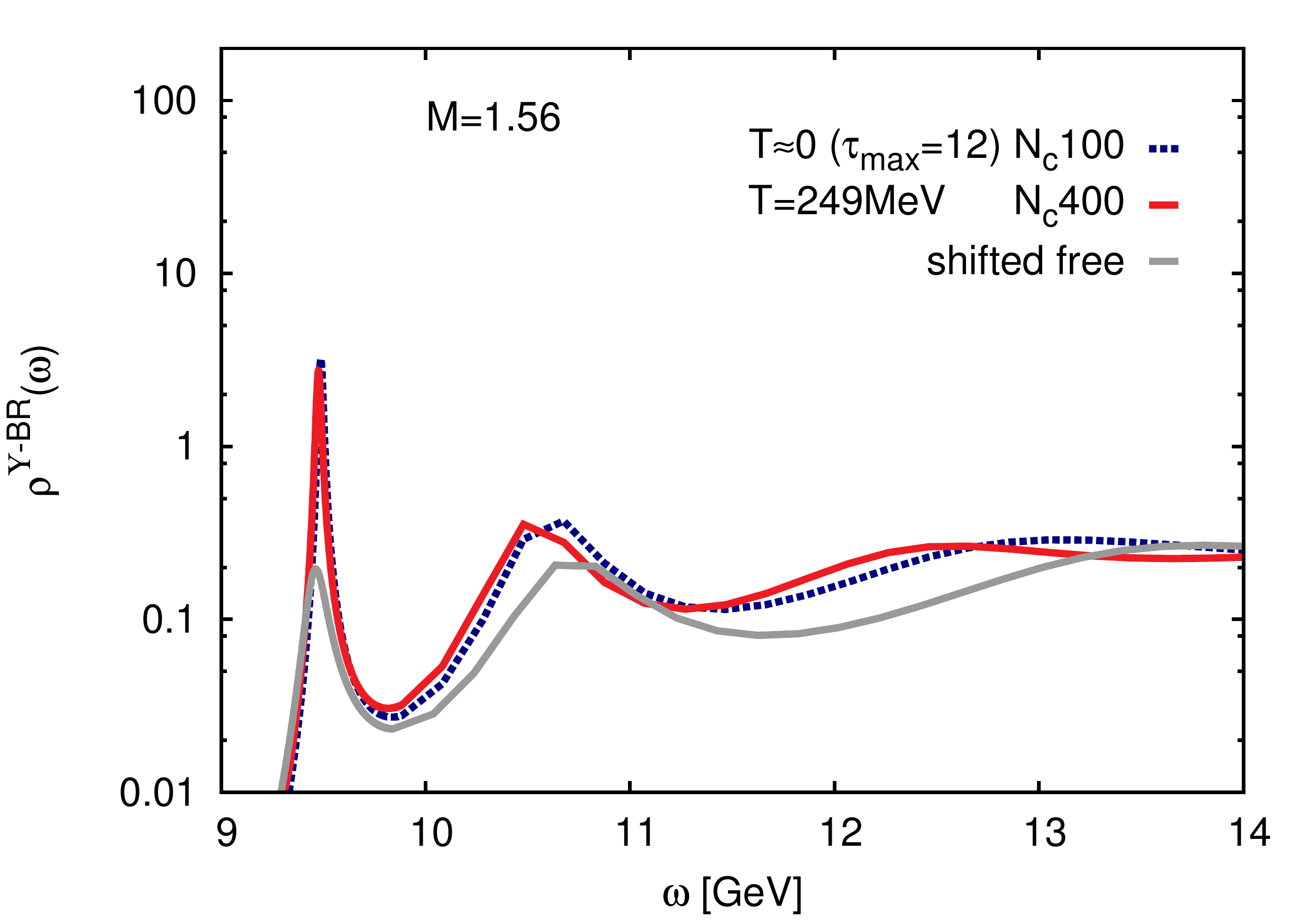}
\includegraphics[scale=0.29,angle=-90]{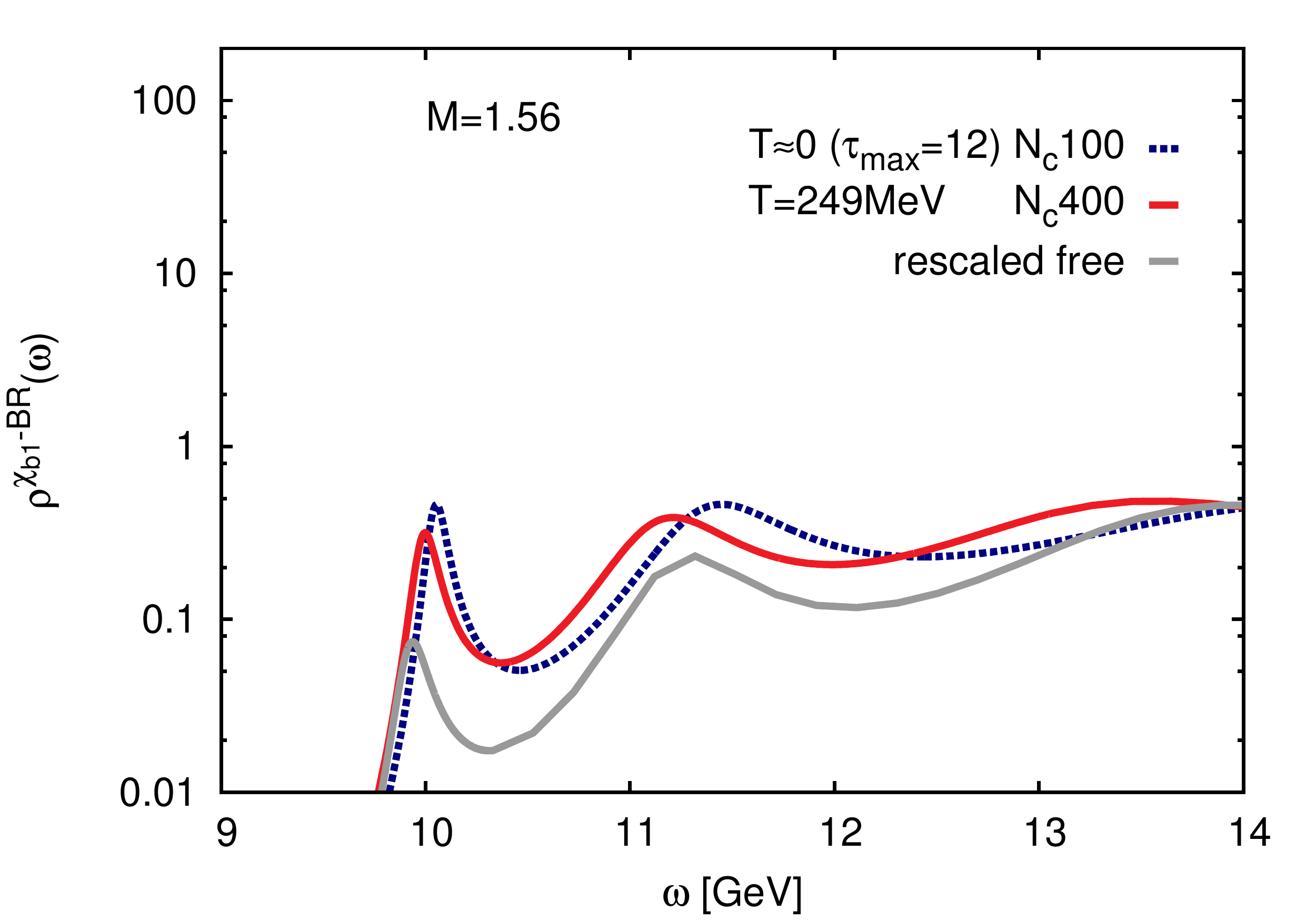}
\end{center}
\caption{The S-wave (left) and P-wave (right) bottomonium spectral functions
at $T=0$ and $T=249$ MeV. The zero temperature spectral functions have been reconstructed
using only the first twelve data points. Also shown are the free spectral functions.}
\label{fig:spf}
\end{figure}

The reconstruction of the spectral function at non-zero temperature
is more difficult. Therefore, as the first step it is worth to
discuss the temperature dependence of the correlation functions.
In Fig. \ref{fig:corr_rat} we show the ratio of the correlators
for S-wave and P-wave bottomonia to the corresponding zero temperature
result. Below or at the transition the changes in the correlators
are quite small, only marginally larger than the statistical errors.
At the two temperatures above the transition we see much larger changes,
although the total size of the temperature dependence does not exceed
$1\%$ for the S-wave correlator and $5\%$ for the P-wave correlator.
The difference of the medium modification of the S-wave and P-wave 
correlators is expected. The larger and more loosely bound $\chi_{b1}$ state
experiences larger in-medium modification than the smaller and more
tightly bound $\Upsilon(1S)$ state.
The reconstruction of the spectral function at non-zero temperature
is more difficult for two reasons. First, the Euclidean time being
limited to $\tau_{max}=1/T$, is short. Second the number of available
data points is also smaller. Therefore when comparing spectral functions
at non-zero temperature with the corresponding zero temperature result
it is important to take into account these systematic effects and distinguish
them from the true medium effects. For this reason we reconstruct the zero
temperature spectral functions for S-wave and P-wave bottomonium using only
the first twelve data points. We then compare the zero temperature spectral functions
reconstructed this way with the finite temperature case, where only twelve data
points are available. This comparison is shown in Fig. \ref{fig:spf} for $\beta=7.28$,
corresponding to $T=249$ MeV. The $omega$-axis in Fig.\ref{fig:spf}  was shifted by $C(\beta)$ defined above.
The ground state peak in the zero temperature spectral functions is significantly
broadened when only the first twelve data points are used in the analysis. 
The difference between the zero temperature and the finite temperature spectral
functions remains small up to this temperature. Furthermore, at low $\omega$
the spectral function are very different from the free spectral functions.
Thus we conclude that 1S and 1P bottomonium states do not melt up
to temperatures $T=249$ MeV.

\begin{figure}
\begin{center}
\centering
\includegraphics[scale=0.29, angle=-90]{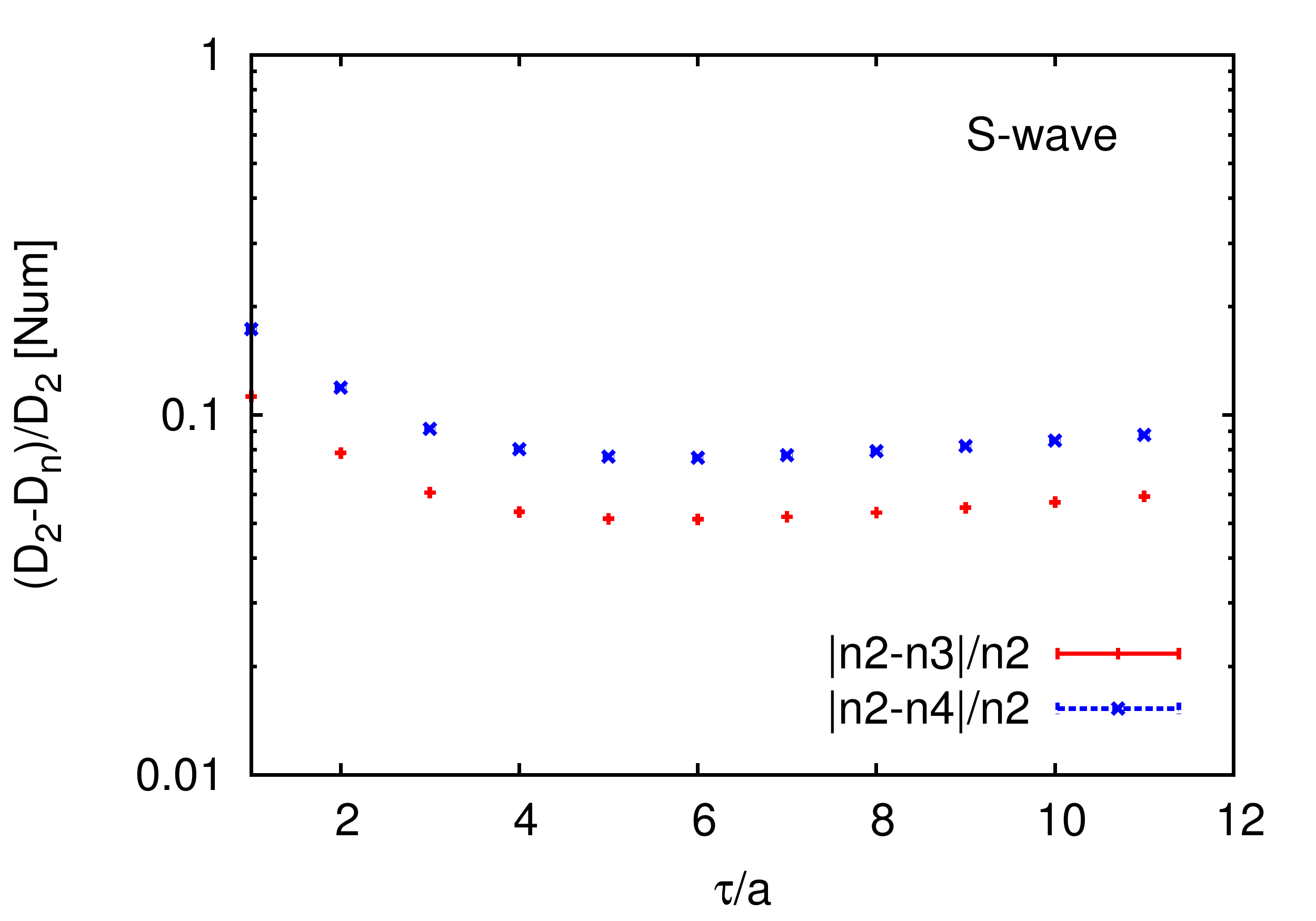}
\includegraphics[scale=0.29, angle=-90]{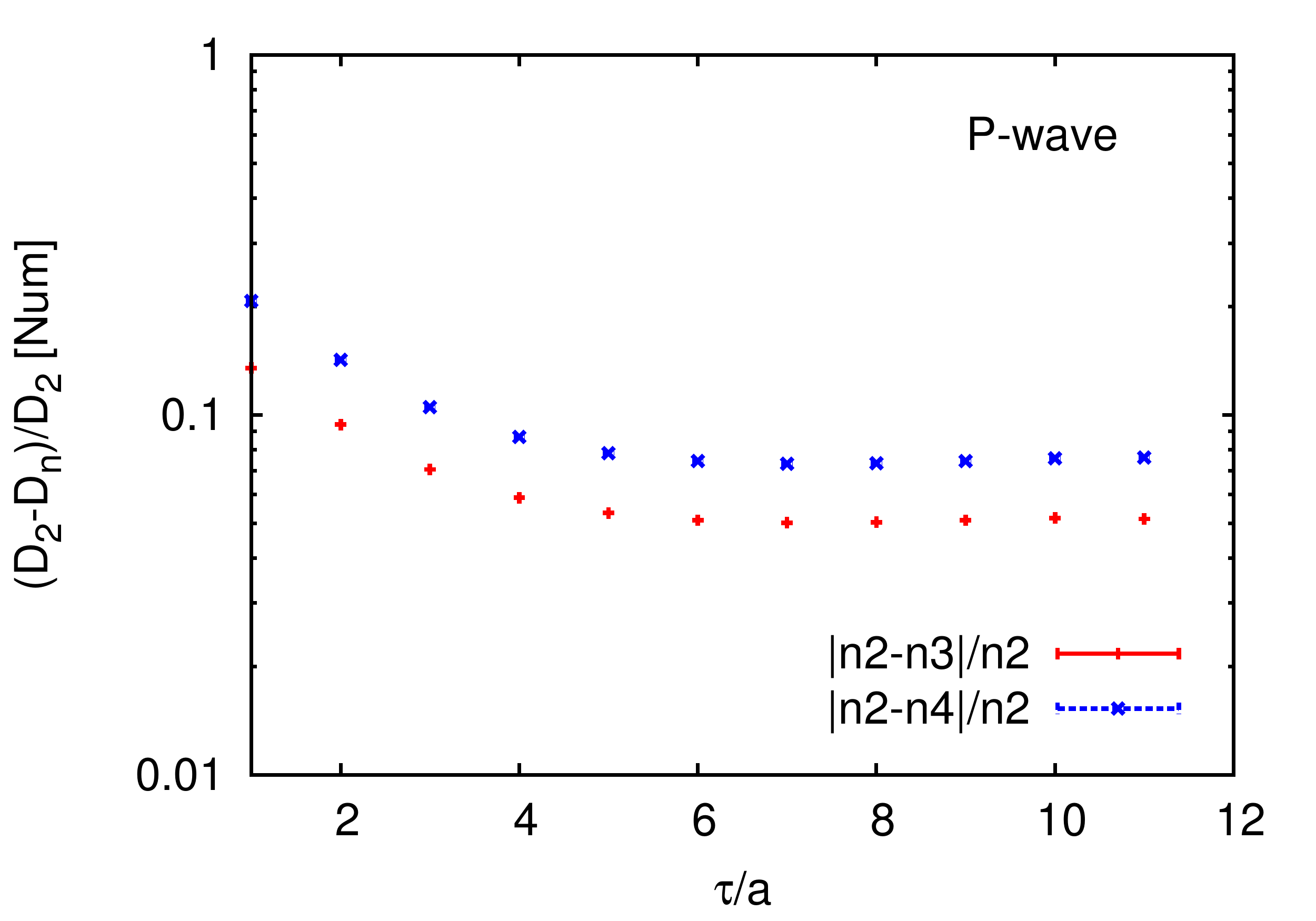}
\end{center}
\caption{The relative difference between correlators calculated with $n=3$ and $4$ and
the correlators calculated with $n=2$ for $T=249$ MeV for S-wave (left) and P-wave (right)
bottomonia.}
\label{fig:diff}
\end{figure}

\begin{figure}
\centering
\includegraphics[scale=0.55]{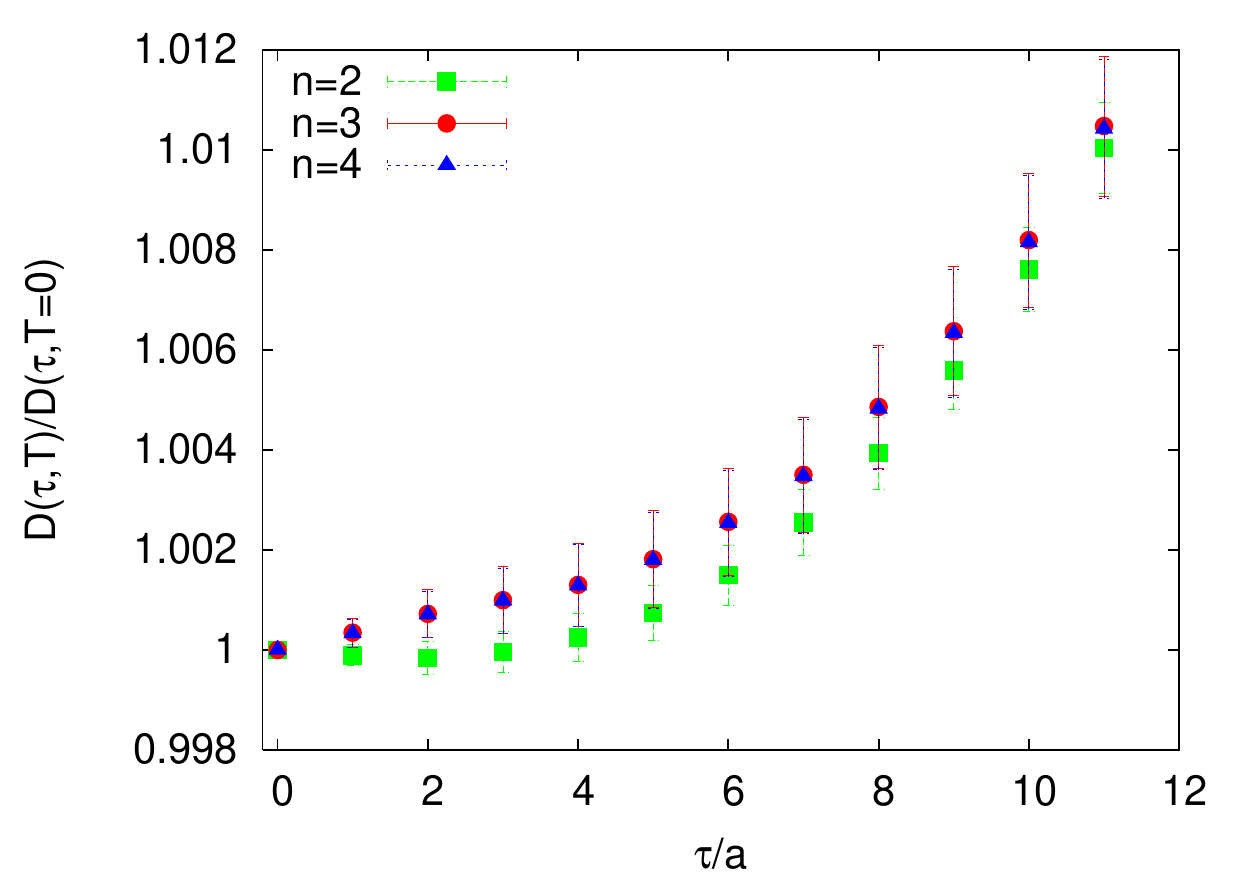}
\includegraphics[scale=0.55]{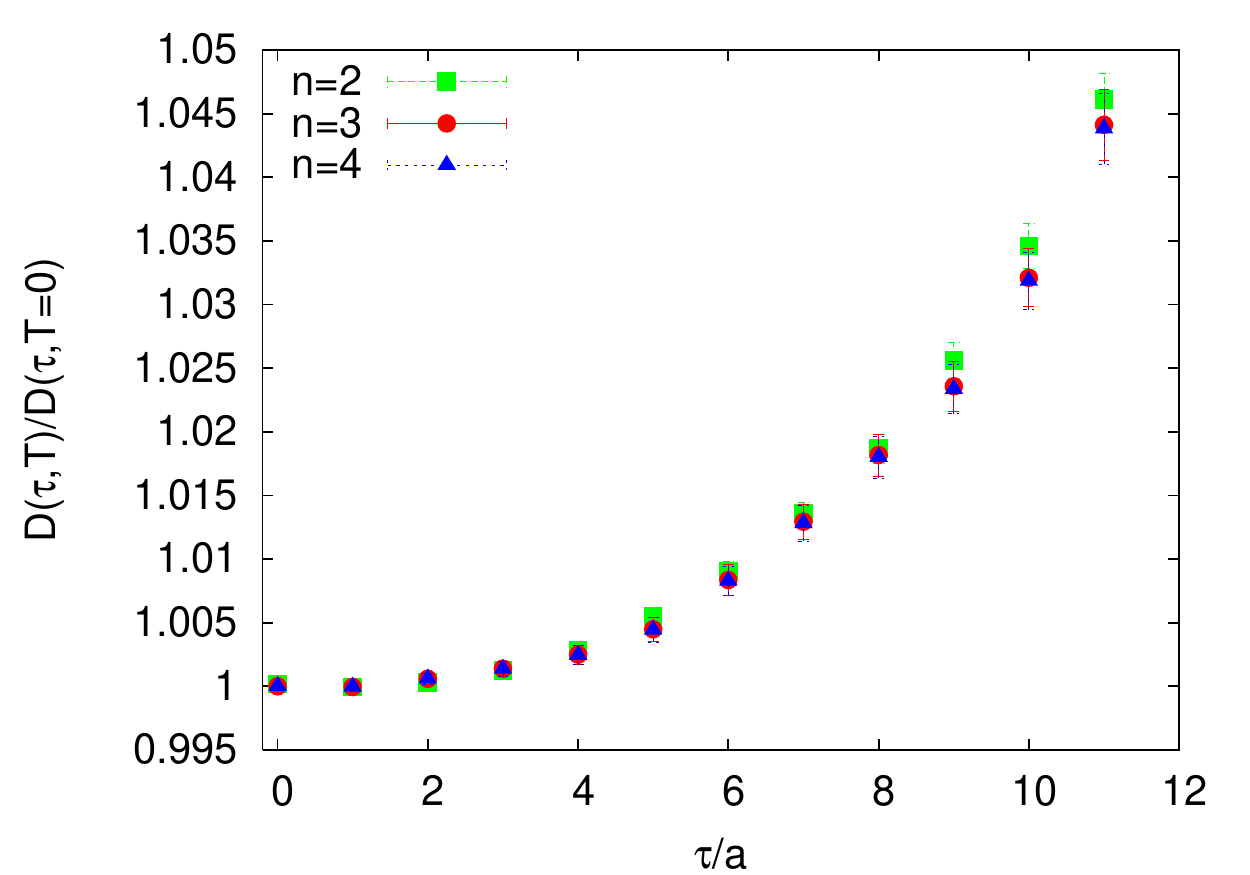}
\caption{The ratio of the S-wave (left) and P-wave (right)  correlators at $T=249$ MeV
to the corresponding zero temperature correlators for the values $n=2,~3$ and $4$ 
of the Lepage parameter.}
\label{fig:rat_ndep}
\end{figure}

So far we presented our numerical results only for $n=2$. It is important to
check if and to what extent our conclusions depend on the temporal discretization
controlled by the Lepage parameter $n$. We performed calculations of bottomonium correlators
also for $n=3$ and $4$. In Fig. \ref{fig:diff} we show the difference
between the bottomonium correlation function calculated with $n=3,~4$ and
$n=2$ for our highest temperature $T=249$ MeV. Since this corresponds also to the finest
lattice used in our study the choice of $n$ is expected to have the largest effect here.
Changing $n$ from $2$ to $3$ and $4$ has the biggest effect at small $\tau$, where it reaches
about $10\%$ for $n=3$ and $20\%$ for $n=4$ for S-wave. For P-wave it can reach $30\%$. For larger $\tau$ the dependence on $n$ becomes smaller.
We observe similar dependence on $n$ in the free theory. In the free theory the dominant
effect of increasing $n$ is the reduction of the high $\omega$ range, where the NRQCD spectral
function has a support \cite{Kim:2014iga}. This is the reason why the effect of changing $n$ is most
prominent at small $\tau$.  
We are mostly interested in the temperature dependence of the bottomonium spectral properties,
and to what extent this temperature dependence is effected by the choice of $n$.
Therefore, we consider the ratio of the bottomonium correlators 
calculated for $T=249$ MeV to the corresponding zero
temperature correlators for $n=2,~3$ and $4$. Our results are shown in Fig. \ref{fig:rat_ndep}
for S-wave and P-wave correlators. 
As one can see from the figure no dependence on $n$ is visible in the ratios 
within the statistical errors, implying that the temperature dependence of the bottomonium
spectral functions is not affected by the choice of $n$.  

\section{Conclusion}
In this contribution we studied bottomonium properties at non-zero temperature using
NQRCD and gauge configurations generated by HotQCD on $48^3 \times 12$ lattices.
We used a novel Bayesian approach to reconstruct the bottomonium spectral functions,
which works reasonably well even for lattices with temporal extent $N_{\tau}=12$.
We see small but statistically significant temperature dependence in the bottomonium
correlators which are larger for the P-wave than for the S-wave, as expected. 
The analysis of the spectral function shows that both $\Upsilon$ and $\chi_{b1}$ states
survive in the deconfined phase up to the highest temperature of $249$ MeV considered
in this study.

\section*{Acknowledgments}
SK is supported
by the National Research Foundation of Korea grant funded by the
Korean government (MEST) No.\ 2010-002219 and in part by
NRF-2008-000458. PP is supported by U.S.Department of Energy under
Contract No.DE-AC02-98CH10886. AR was partly supported by the Swiss
National Science Foundation (SNF) under grant 200021-140234.
This paper was finished during our stay at the Institute for Nuclear Theory (INT) 
program Heavy Flavor and Electromagnetic Probes in Heavy Ion Collisions (INT-14-3).
We thank INT for support.

\end{document}